\documentclass[twocolumn, superscriptaddress, prl, nobibnotes]{revtex4-2}

\usepackage{amsmath,amssymb}
\usepackage{natbib}
\usepackage{etoolbox}
\setcitestyle{super}
\makeatletter
\newcommand*{\citenumns}[2][]{%
  \begingroup
  \let\NAT@mbox=\mbox
  \let\@cite\NAT@citenum
  \let\NAT@space\NAT@spacechar
  \let\NAT@super@kern\relax
  \renewcommand\NAT@open{}%
  \renewcommand\NAT@close{}%
  \cite[#1]{#2}%
  \endgroup
}
\makeatother
\patchcmd{\section}
  {\centering}
  {\raggedright}
  {}
  {}
\patchcmd{\subsection}
  {\centering}
  {\raggedright}
  {}
  {}

\usepackage{xcolor}
\usepackage{graphicx}

\usepackage[T1]{fontenc}
\usepackage[colorlinks,hyperindex]{hyperref}
\hypersetup
{
   colorlinks,%
   citecolor=blue,%
   linkcolor=blue,%
   urlcolor=blue,%
}
\usepackage[capitalise]{cleveref}
\usepackage{orcidlink}

\newcommand\gs{\mathrm{g.s.}}
\newcommand\logft{\mathrm{log}ft}
\newcommand\bgt{B(\mathrm{GT})}

\makeindex

\begin{document}

\title{Beta decay of potassium-54 reveals the role of\\two-body currents in neutron-rich nuclei}
\author{Z.~Y.~Xu\orcidlink{0000-0001-8626-1276}}
\affiliation{Department of Physics and Astronomy, University of Tennessee, Knoxville, TN 37996, USA}
\author{R.~Grzywacz\orcidlink{0000-0002-0920-2587}}
\affiliation{Department of Physics and Astronomy, University of Tennessee, Knoxville, TN 37996, USA}
\author{M.~Heinz\orcidlink{0000-0002-6363-0056}}
\affiliation{National Center for Computational Sciences, Oak Ridge National Laboratory, Oak Ridge, TN 37831, USA}
\affiliation{Physics Division, Oak Ridge National Laboratory, Oak Ridge, TN 37831, USA}
\author{T.~Papenbrock\orcidlink{0000-0001-8733-2849}}
\affiliation{Department of Physics and Astronomy, University of Tennessee, Knoxville, TN 37996, USA}
\affiliation{Physics Division, Oak Ridge National Laboratory, Oak Ridge, TN 37831, USA}
\author{W.-J.~Ong\orcidlink{0000-0002-0945-8654}}
\affiliation{Lawrence Livermore National Laboratory, Livermore, CA 94550, USA}
\author{J.~M.~Allmond\orcidlink{0000-0001-6533-8721}}
\affiliation{Physics Division, Oak Ridge National Laboratory, Oak Ridge, TN 37831, USA}
\author{B.~C.~Rasco\orcidlink{0000-0002-4061-1178}}
\affiliation{Physics Division, Oak Ridge National Laboratory, Oak Ridge, TN 37831, USA}
\author{K.~P.~Rykaczewski\orcidlink{0000-0002-5717-3418}}
\affiliation{Physics Division, Oak Ridge National Laboratory, Oak Ridge, TN 37831, USA}
\author{H.~Schatz\orcidlink{0000-0003-1674-4859}}
\affiliation{Department of Physics and Astronomy, Michigan State University, East Lansing, MI 48824, USA}
\affiliation{Facility for Rare Isotope Beams, Michigan State University, East Lansing, MI 48824, USA}
\author{B.~M.~Sherrill\orcidlink{0000-0002-5093-8190}}
\affiliation{Department of Physics and Astronomy, Michigan State University, East Lansing, MI 48824, USA}
\affiliation{Facility for Rare Isotope Beams, Michigan State University, East Lansing, MI 48824, USA}
\author{O.~B.~Tarasov\orcidlink{0000-0003-0497-1421}}
\affiliation{Facility for Rare Isotope Beams, Michigan State University, East Lansing, MI 48824, USA}
\author{P.~Gysbers\orcidlink{0000-0002-5269-9170}}
\affiliation{Facility for Rare Isotope Beams, Michigan State University, East Lansing, MI 48824, USA}
\author{G.~Hagen\orcidlink{0000-0001-6019-1687}}
\affiliation{Physics Division, Oak Ridge National Laboratory, Oak Ridge, TN 37831, USA}
\affiliation{Department of Physics and Astronomy, University of Tennessee, Knoxville, TN 37996, USA}
\author{P.~Navr{\'a}til\orcidlink{0000-0002-6535-2141}}
\affiliation{TRIUMF, Vancouver, British Columbia V6T 2A3, Canada}
\affiliation{University of Victoria, Victoria, British Columbia V8P 5C2, Canada}
\author{S.~Ajayi\orcidlink{0000-0002-3564-4856}}
\affiliation{Department of Physics, Florida State University, Tallahassee, FL 32306, USA}
\author{H.~Arora\orcidlink{0000-0001-6183-658X}}
\affiliation{Department of Physics, Central Michigan University, Mt Pleasant, MI 48858, USA}
\affiliation{Center for Exotic Nuclear Studies, Institute for Basic Science, Daejeon 34126, Republic of Korea}
\author{A.~D.~Ayangeakaa\orcidlink{0000-0003-1679-3175}}
\affiliation{Department of Physics and Astronomy, University of North Carolina Chapel Hill, NC 27599, USA}
\affiliation{Triangle Universities Nuclear Laboratory, Duke University, Durham, NC 27708, USA}
\author{H.~C.~Berg\orcidlink{0000-0003-2090-1255}}
\affiliation{Department of Physics and Astronomy, Michigan State University, East Lansing, MI 48824, USA}
\affiliation{Facility for Rare Isotope Beams, Michigan State University, East Lansing, MI 48824, USA}
\author{J.~M.~Berkman\orcidlink{0009-0001-9753-0096}}
\affiliation{Department of Chemistry, Michigan State University, East Lansing, MI 48824, USA}
\affiliation{Facility for Rare Isotope Beams, Michigan State University, East Lansing, MI 48824, USA}
\author{D.~L.~Bleuel\orcidlink{0000-0003-2792-0715}}
\affiliation{Lawrence Livermore National Laboratory, Livermore, CA 94550, USA}
\author{K.~Bosmpotinis\orcidlink{0000-0002-9807-0001}}
\affiliation{Department of Physics and Astronomy, Michigan State University, East Lansing, MI 48824, USA}
\affiliation{Facility for Rare Isotope Beams, Michigan State University, East Lansing, MI 48824, USA}
\author{N.~Braukman\orcidlink{0009-0000-4666-0440}}
\affiliation{Department of Physics and Astronomy, University of Tennessee, Knoxville, TN 37996, USA}
\author{M.~P.~Carpenter\orcidlink{0000-0002-3237-5734}}
\affiliation{Physics Division, Argonne National Laboratory, Argonne, IL 60439, USA}
\author{G.~Cerizza\orcidlink{0009-0005-1063-9996}}
\affiliation{Facility for Rare Isotope Beams, Michigan State University, East Lansing, MI 48824, USA}
\author{A.~Chester\orcidlink{0000-0002-7625-7772}}
\affiliation{Facility for Rare Isotope Beams, Michigan State University, East Lansing, MI 48824, USA}
\author{J.~M.~Christie\orcidlink{0000-0002-8653-8236}}
\affiliation{Department of Physics and Astronomy, University of Tennessee, Knoxville, TN 37996, USA}
\affiliation{Physics Division, Oak Ridge National Laboratory, Oak Ridge, TN 37831, USA}
\author{I.~Cox\orcidlink{0000-0002-6807-3658}}
\affiliation{Department of Physics and Astronomy, University of Tennessee, Knoxville, TN 37996, USA}
\author{H.~L.~Crawford\orcidlink{0000-0002-7765-4235}}
\affiliation{Nuclear Science Division, Lawrence Berkeley National Laboratory, Berkeley, CA 94720, USA}
\author{B.~P.~Crider\orcidlink{0000-0001-7153-8762}}
\affiliation{Department of Physics and Astronomy, Mississippi State University, Mississippi State, MS 39762, USA}
\author{J.~Davis\orcidlink{0009-0007-0135-2197}}
\affiliation{Pacific Northwest National Laboratory, Richland, WA 99354, USA}
\affiliation{Department of Physics, Florida State University, Tallahassee, FL 32306, USA}
\author{A.~A.~Doetsch\orcidlink{0000-0001-7223-789X}}
\affiliation{Department of Physics and Astronomy, Michigan State University, East Lansing, MI 48824, USA}
\affiliation{Facility for Rare Isotope Beams, Michigan State University, East Lansing, MI 48824, USA}
\author{J.~G.~Duarte\orcidlink{0000-0002-8781-3834}}
\affiliation{Lawrence Livermore National Laboratory, Livermore, CA 94550, USA}
\author{A.~Estrade\orcidlink{0000-0002-0537-9043}}
\affiliation{Department of Physics, Central Michigan University, Mt Pleasant, MI 48858, USA}
\author{A.~Fija\l{}kowska\orcidlink{0000-0001-6472-4284}}
\affiliation{Faculty of Physics, University of Warsaw, Warsaw, PL 02-093, Poland}
\author{C.~Frantzis\orcidlink{0009-0000-1482-5933}}
\affiliation{Department of Physics and Astronomy, Michigan State University, East Lansing, MI 48824, USA}
\affiliation{Facility for Rare Isotope Beams, Michigan State University, East Lansing, MI 48824, USA}
\author{K.~Fukushima\orcidlink{0000-0002-7124-7591}}
\affiliation{Facility for Rare Isotope Beams, Michigan State University, East Lansing, MI 48824, USA}
\author{T.~Gaballah\orcidlink{0009-0003-2846-5117}}
\affiliation{Department of Physics and Astronomy, Mississippi State University, Mississippi State, MS 39762, USA}
\affiliation{Facility for Rare Isotope Beams, Michigan State University, East Lansing, MI 48824, USA}
\author{T.~J.~Gray\orcidlink{0000-0003-3965-6130}}
\affiliation{Department of Physics and Astronomy, University of Tennessee, Knoxville, TN 37996, USA}
\affiliation{Physics Division, Oak Ridge National Laboratory, Oak Ridge, TN 37831, USA}
\author{E.~C.~Good\orcidlink{0000-0001-7765-6188}}
\affiliation{Facility for Rare Isotope Beams, Michigan State University, East Lansing, MI 48824, USA}
\author{K.~Haak\orcidlink{0009-0004-2272-4250}}
\affiliation{Department of Physics and Astronomy, Michigan State University, East Lansing, MI 48824, USA}
\affiliation{Facility for Rare Isotope Beams, Michigan State University, East Lansing, MI 48824, USA}
\author{S.~Hanai}
\affiliation{Center for Nuclear Study, University of Tokyo, Wako, Saitama 351-0198, Japan}
\author{J.~T.~Harke\orcidlink{0000-0001-5130-7123}}
\affiliation{Lawrence Livermore National Laboratory, Livermore, CA 94550, USA}
\author{C.~M.~Harris\orcidlink{0000-0002-4101-6098}}
\affiliation{Department of Physics and Astronomy, Michigan State University, East Lansing, MI 48824, USA}
\affiliation{Facility for Rare Isotope Beams, Michigan State University, East Lansing, MI 48824, USA}
\author{A.~C.~Hartley}
\affiliation{Department of Physics and Astronomy, Michigan State University, East Lansing, MI 48824, USA}
\affiliation{Facility for Rare Isotope Beams, Michigan State University, East Lansing, MI 48824, USA}
\author{M.~Hausmann\orcidlink{0000-0002-1166-8378}}
\affiliation{Facility for Rare Isotope Beams, Michigan State University, East Lansing, MI 48824, USA}
\author{K.~Hermansen\orcidlink{0000-0002-1707-0844}}
\affiliation{Department of Physics and Astronomy, Michigan State University, East Lansing, MI 48824, USA}
\affiliation{Facility for Rare Isotope Beams, Michigan State University, East Lansing, MI 48824, USA}
\author{D.~E.~M.~Hoff\orcidlink{0000-0003-0838-5892}}
\affiliation{Lawrence Livermore National Laboratory, Livermore, CA 94550, USA}
\author{D.~Hoskins III\orcidlink{0009-0005-9127-8140}}
\affiliation{Department of Physics and Astronomy, University of Tennessee, Knoxville, TN 37996, USA}
\author{J.~Huffman\orcidlink{0009-0000-1536-1624}}
\affiliation{Department of Physics and Astronomy, Michigan State University, East Lansing, MI 48824, USA}
\affiliation{Facility for Rare Isotope Beams, Michigan State University, East Lansing, MI 48824, USA}
\author{R.~Jain\orcidlink{0000-0001-9859-1512}}
\affiliation{Lawrence Livermore National Laboratory, Livermore, CA 94550, USA}
\affiliation{Department of Physics and Astronomy, Michigan State University, East Lansing, MI 48824, USA}
\affiliation{Facility for Rare Isotope Beams, Michigan State University, East Lansing, MI 48824, USA}
\author{M.~Karny\orcidlink{0000-0002-1050-6203}}
\affiliation{Faculty of Physics, University of Warsaw, Warsaw, PL 02-093, Poland}
\author{T.~T.~King\orcidlink{0000-0003-0649-3275}}
\affiliation{Physics Division, Oak Ridge National Laboratory, Oak Ridge, TN 37831, USA}
\author{N.~Kitamura}
\affiliation{Center for Nuclear Study, University of Tokyo, Wako, Saitama 351-0198, Japan}
\affiliation{Department of Physics and Astronomy, University of Tennessee, Knoxville, TN 37996, USA}
\author{K.~Kolos\orcidlink{0000-0002-1726-4171}}
\affiliation{Lawrence Livermore National Laboratory, Livermore, CA 94550, USA}
\author{E.~Kwan}
\affiliation{Facility for Rare Isotope Beams, Michigan State University, East Lansing, MI 48824, USA}
\author{S.~N.~Liddick\orcidlink{0000-0003-1589-891X}}
\affiliation{Department of Chemistry, Michigan State University, East Lansing, MI 48824, USA}
\affiliation{Facility for Rare Isotope Beams, Michigan State University, East Lansing, MI 48824, USA}
\author{B.~Longfellow\orcidlink{0000-0001-8743-962X}}
\affiliation{Lawrence Livermore National Laboratory, Livermore, CA 94550, USA}
\author{R.~S.~Lubna\orcidlink{0000-0003-4976-3976}}
\affiliation{Facility for Rare Isotope Beams, Michigan State University, East Lansing, MI 48824, USA}
\author{S.~Lyons\orcidlink{0000-0002-0194-261X}}
\affiliation{Pacific Northwest National Laboratory, Richland, WA 99354, USA}
\author{M.~Madurga\orcidlink{0000-0002-8177-4328}}
\affiliation{Department of Physics and Astronomy, University of Tennessee, Knoxville, TN 37996, USA}
\author{M.~J.~Mogannam\orcidlink{0000-0001-8951-4953}}
\affiliation{Department of Chemistry, Michigan State University, East Lansing, MI 48824, USA}
\affiliation{Facility for Rare Isotope Beams, Michigan State University, East Lansing, MI 48824, USA}
\author{S.~Neupane\orcidlink{0000-0003-3339-8353}}
\affiliation{Lawrence Livermore National Laboratory, Livermore, CA 94550, USA}
\affiliation{Department of Physics and Astronomy, University of Tennessee, Knoxville, TN 37996, USA}
\author{T.~H.~Ogunbeku\orcidlink{0000-0002-4442-7757}}
\affiliation{Lawrence Livermore National Laboratory, Livermore, CA 94550, USA}
\affiliation{Department of Physics and Astronomy, Mississippi State University, Mississippi State, MS 39762, USA}
\affiliation{Facility for Rare Isotope Beams, Michigan State University, East Lansing, MI 48824, USA}
\author{G.~Owens-Fryar\orcidlink{0000-0001-6728-6494}}
\affiliation{Department of Physics and Astronomy, Michigan State University, East Lansing, MI 48824, USA}
\affiliation{Facility for Rare Isotope Beams, Michigan State University, East Lansing, MI 48824, USA}
\author{J.~R.~Palomino\orcidlink{0009-0003-7194-3663}}
\affiliation{Department of Physics and Astronomy, Mississippi State University, Mississippi State, MS 39762, USA}
\author{M.~Portillo\orcidlink{0000-0002-9070-1841}}
\affiliation{Facility for Rare Isotope Beams, Michigan State University, East Lansing, MI 48824, USA}
\author{M.~M.~Rajabali\orcidlink{0000-0002-1046-4998}}
\affiliation{Physics Department, Tennessee Technological University, Cookeville, TN 38505, USA}
\author{A.~L.~Richard\orcidlink{0000-0001-8308-688X}}
\affiliation{Lawrence Livermore National Laboratory, Livermore, CA 94550, USA}
\affiliation{Department of Physics and Astronomy, Ohio University, Athens, OH 45701, USA}
\author{I.~Richardson\orcidlink{0009-0005-3843-6058}}
\affiliation{Department of Physics and Astronomy, Michigan State University, East Lansing, MI 48824, USA}
\affiliation{Facility for Rare Isotope Beams, Michigan State University, East Lansing, MI 48824, USA}
\author{E.~K.~Ronning\orcidlink{0009-0006-4672-5834}}
\affiliation{Department of Chemistry, Michigan State University, East Lansing, MI 48824, USA}
\affiliation{Facility for Rare Isotope Beams, Michigan State University, East Lansing, MI 48824, USA}
\affiliation{Istituto Nazionale di Fisica Nucleare, Sezione di Padova, I-35131 Padova, Italy}
\author{G.~E.~Rose\orcidlink{0000-0002-5645-8905}}
\affiliation{University of California Berkeley, Berkeley, CA 94704, USA}
\author{T.~Ruland\orcidlink{0000-0003-3354-1754}}
\affiliation{Physics Division, Oak Ridge National Laboratory, Oak Ridge, TN 37831, USA}
\author{N.~D.~Scielzo\orcidlink{0000-0003-4976-6552}}
\affiliation{Lawrence Livermore National Laboratory, Livermore, CA 94550, USA}
\author{D.~P.~Scriven\orcidlink{0000-0002-3853-2200}}
\affiliation{Facility for Rare Isotope Beams, Michigan State University, East Lansing, MI 48824, USA}
\author{D.~Seweryniak\orcidlink{0009-0003-3028-9693}}
\affiliation{Physics Division, Argonne National Laboratory, Argonne, IL 60439, USA}
\author{K.~Siegl\orcidlink{0000-0002-7778-315X}}
\affiliation{Department of Physics and Astronomy, University of Tennessee, Knoxville, TN 37996, USA}
\author{M.~K.~Smith\orcidlink{0009-0009-3030-8341}}
\affiliation{Facility for Rare Isotope Beams, Michigan State University, East Lansing, MI 48824, USA}
\author{A.~Spyrou\orcidlink{0000-0002-8642-8352}}
\affiliation{Department of Physics and Astronomy, Michigan State University, East Lansing, MI 48824, USA}
\affiliation{Facility for Rare Isotope Beams, Michigan State University, East Lansing, MI 48824, USA}
\author{M.~Stepaniuk\orcidlink{0000-0002-9631-1713}}
\affiliation{Faculty of Physics, University of Warsaw, Warsaw, PL 02-093, Poland}
\affiliation{Department of Physics and Astronomy, Louisiana State University, Baton Rouge, LA 70803 USA}
\author{A.~Sweet\orcidlink{0000-0001-7750-5232}}
\affiliation{Lawrence Livermore National Laboratory, Livermore, CA 94550, USA}
\author{V.~Tripathi\orcidlink{0000-0002-8850-1357}}
\affiliation{Department of Physics, Florida State University, Tallahassee, FL 32306, USA}
\author{A.~Tsantiri\orcidlink{0000-0002-9759-3476}}
\affiliation{Department of Physics and Astronomy, Michigan State University, East Lansing, MI 48824, USA}
\affiliation{Facility for Rare Isotope Beams, Michigan State University, East Lansing, MI 48824, USA}
\affiliation{University of Regina, 3737 Wascana Parkway, Regina, Saskatchewan S4S 0A2, Canada}
\author{S.~Uthayakumaar\orcidlink{0000-0002-3154-0734}}
\affiliation{Facility for Rare Isotope Beams, Michigan State University, East Lansing, MI 48824, USA}
\author{W.~B.~Walters\orcidlink{0000-0001-5692-6943}}
\affiliation{Department of Chemistry and Biochemistry, University of Maryland, College Park, MD 20742, USA}
\author{S.~Watters\orcidlink{0000-0002-6243-394X}}
\affiliation{Department of Physics and Astronomy, Michigan State University, East Lansing, MI 48824, USA}
\affiliation{Facility for Rare Isotope Beams, Michigan State University, East Lansing, MI 48824, USA}
\author{M.~Wolinska-Cichocka\orcidlink{0000-0002-7857-7796}}
\affiliation{Heavy Ion Laboratory, University of Warsaw, PL-02-093 Warsaw, Poland}
\author{R.~Yokoyama\orcidlink{0000-0003-3298-3542}}
\affiliation{Center for Nuclear Study, University of Tokyo, Wako, Saitama 351-0198, Japan}

\maketitle
\onecolumngrid
\noindent $\ast$Corresponding Author: Zhengyu Xu (zxu41@utk.edu).\\
$\ast$Corresponding Author: Robert Grzywacz (rgrzywac@utk.edu).\\
$\ast$Corresponding Author: Matthias Heinz (heinzmc@ornl.gov).

\section*{Abstract}
The $\beta$ decay of neutron-rich nuclei probes the microscopic nature of
weak interactions in the nuclear medium under extreme conditions. Such weak
decays drive the synthesis of heavy elements in astrophysical environments,
but remain poorly understood. Far from stability, experimental data are
scarce, placing strong demands on ab~initio nuclear theory. Two-body weak
currents have been proposed as the dominant mechanism responsible for the
universal quenching of theoretically predicted Gamow-Teller decay
strength. However, direct experimental validation in neutron-rich nuclei has
been lacking. Here we measure the $\beta$ decay of $^{54}$K, a nucleus at the
experimental frontier of radioactive ion-beam facilities. In this work, the
doubly magic nature of the daughter nucleus $^{54}$Ca results in a remarkably
simple decay pattern, dominated by a single transition to a neutron-unbound
state at 6.2 MeV with a comparative partial half-life $\logft=4.4(1)$ or
Gamow-Teller decay strength $\bgt=0.26^{+0.08}_{-0.04}$. Our measurement is
reproduced by new state-of-the-art calculations of the decay, providing
direct evidence for the essential role of two-body currents in nuclear
$\beta$ decay for systems with a significant neutron excess. We establish a
key benchmark for theoretical predictions of weak processes in exotic nuclei.\\

\twocolumngrid
\section*{Introduction}
Nuclear $\beta$ decay, a weak-interaction process that converts a neutron to a
proton under the emission of two leptons, an electron and an electron
antineutrino, is one of the most common types of radioactivity that transforms
neutron-rich nuclei. It is an effective probe to study nuclear structure far
from stability as well as physics beyond the standard model~\cite{Pfut2012,
Rubi2025}. Furthermore, $\beta$ decay plays a pivotal role in various
nucleosynthesis processes in the Universe by guiding the nuclear mass flow to
the neighboring isotopic or isotonic chains~\cite{Cowa2021, Arno2007, Wied2025}.

\begin{figure}[ht]
   \centering
   \includegraphics[width=0.49\textwidth]{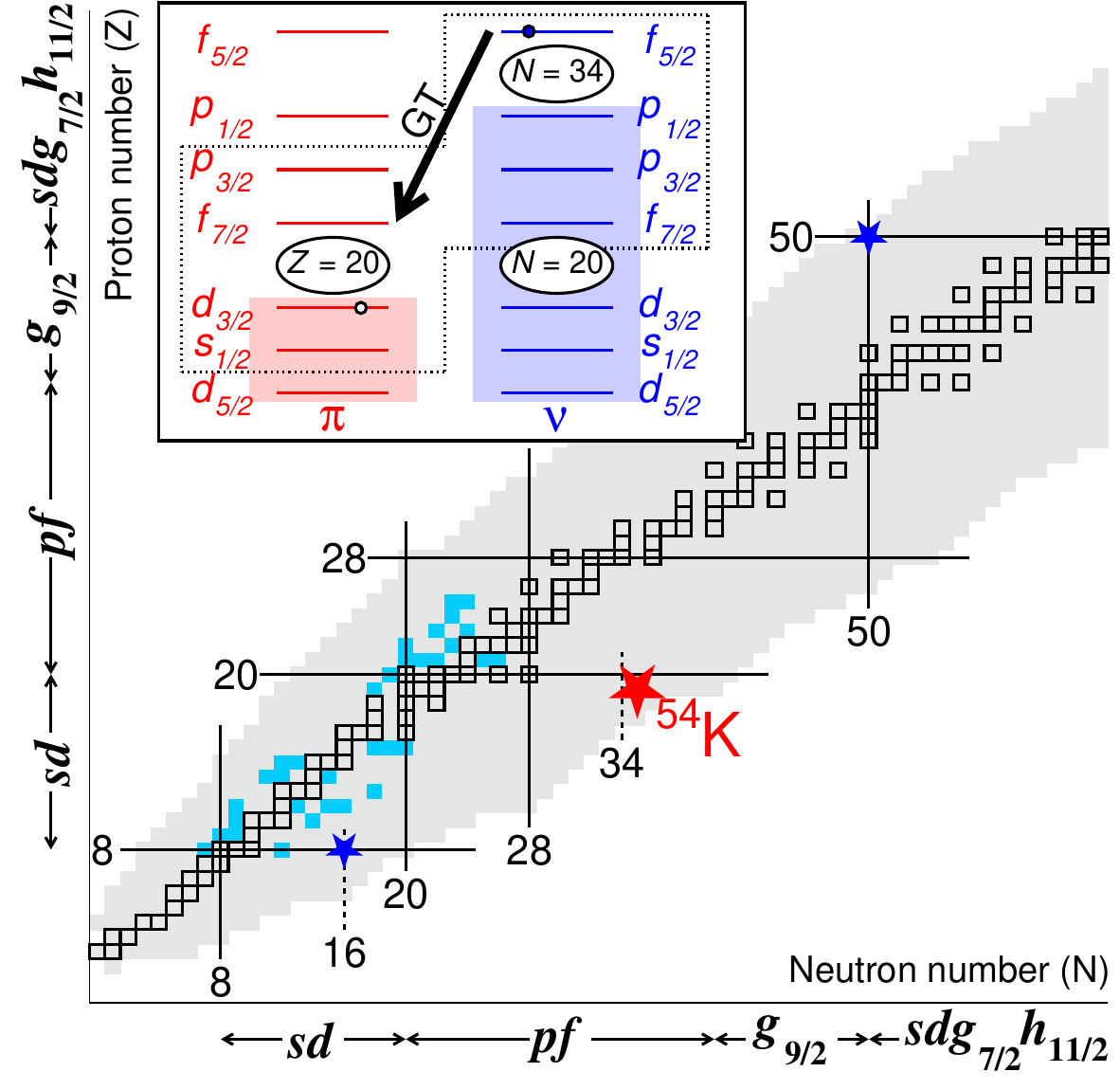}
   \caption{
      The chart of nuclei highlighting the medium-mass isotopes where
      Gamow-Teller strengths have been predicted by ab~initio methods including
      2BCs~\cite{Gysb2019, Stro2021, Neup2024}, with the doubly magic $^{24}$O
      and $^{100}$Sn drawn as blue stars and the rest as turquoise squares.
      Stable nuclei are shown with black outlines. Horizontal and vertical lines
      indicate the proton and neutron magic numbers, respectively, in the shell
      model, and the labels $sd$, $pf$, etc.\ along the axes refer to the
      single-particle orbitals in the corresponding harmonic-oscillator
      shells~\cite{Maye1955}. The short, vertical dashed lines are drawn for
      neutron magic numbers 16 and 34 which only prevail in the neutron-rich
      nuclei. The inset displays a schematic diagram of single-particle orbitals
      in the $sd$-$pf$ shells, with the proton-neutron occupation in the
      $^{54}$K ground state. The last neutron particle (proton hole) at the
      Fermi surface is represented by the solid (open) circle on the $\nu
      f_{5/2}$ ($\pi d_{3/2}$) orbital. The dotted lines enclose the valence
      space for our ab~initio calculations. The diagonal arrow represents the
      strong Gamow-Teller channel investigated in this work.
   }\label{chart}
\end{figure}

In the allowed Gamow-Teller $\beta$ decay (since we exclusively deal with
allowed $\beta$ decay in this article, ``allowed'' will be omitted hereafter for
brevity), in which the two leptons are emitted with orbital angular momentum
$L=0$ and spin $S=1$~\cite{Gamo1936}, there is a long-standing puzzle about why
the decay rates in nuclei are lower than what is expected from the free neutron
decay~\cite{Town1987}, creating the appearance of a quenching factor $q\sim0.75$
on the axial-vector coupling, $g_A=1.27$, in theory calculations~\cite{Chou1993,
Mart1996}. Ab~initio computations have shown that the apparent quenching is due
to strong interaction correlations in the nucleus and crucially two-body
currents (2BCs), an in-medium effect modifying the Gamow-Teller operator such
that it couples to two interacting nucleons~\cite{Mene2011, Gysb2019}. This
result is the third major milestone in significantly advancing our comprehension
of $\beta$ decay since the formulation of Fermi’s theory~\cite{Ferm1934} and the
understanding of the role of configuration interactions~\cite{Bell1958}.
Figure~\ref{chart} highlights medium-mass nuclei where measured Gamow-Teller
decays have confronted computations including 2BCs to date~\cite{Gysb2019,
Stro2021, Neup2024}: Most are close to $\beta$-stability (shown as black
squares), with the exception of the neutron-deficient $^{100}$Sn (with $Z=50$
protons and $N=50$ neutrons) and the neutron-rich $^{24}$O ($Z=8$ and $N=16$).
For the latter, the transition matrix of the Gamow-Teller decay is somewhat
small~\cite{Neup2024}, making the 2BCs effect less prominent and their role more
complicated to isolate.

In this work, we measured the Gamow-Teller strength of $^{54}$K ($Z=19$ and
$N=35$). Comparing with new state-of-the-art ab~initio calculations, we validate
the quenching induced by 2BCs in an extremely neutron-rich nucleus. At the
experimental frontier of current radioactive ion-beam facilities, this
transition, with its relatively large matrix element and significant cross-shell
excitations, is a key benchmark of ab~initio predictions of $\beta$ decays in
the exotic nuclei with substantially more neutrons than protons, allowing us to
extend our model to experimentally inaccessible nuclei that are produced or
involved in the nucleosynthetic flow of astrophysical events~\cite{Thie2017}.

In the nuclear shell model~\cite{Maye1955}, $^{54}$K can be viewed as a neutron
particle and a proton hole with respect to the doubly magic $^{54}$Ca ($Z=20$
and $N=34$), in which the neutron magicity is induced by the energy gap between
the $\nu p_{1/2}$ (we use $\nu$ and $\pi$ for neutron and proton orbitals,
respectively) and $\nu f_{5/2}$ orbitals ($\Delta E\sim2.2$ MeV~\cite{Wang2021})
at the $Z=20$ proton shell closure~\cite{Step2013, Mich2018, Chen2019,
Ogun2025}. Experimental results reveal that the $N=34$ shell closure disappears
rapidly above $Z=20$, i.e., in the scandium ($Z=21$), titanium ($Z=22$), and
vanadium ($Z=23$) isotopes~\cite{Leis2021, Port2022, Iimu2023}, but persists in
calcium and below~\cite{Step2013, Mich2018, Chen2019, Liu2019, Ogun2025,
Zeng2025}.

Following such single-particle scheme, see the inset of \cref{chart}, the last
neutron in $^{54}$K occupies the $\nu f_{5/2}$ orbital above the $N=34$ shell
gap. With a fully open proton $pf$ shell outside the $Z=20$ core, a strong
Gamow-Teller transition $\nu f_{5/2}\rightarrow\pi f_{7/2}$ is anticipated to
dominate its $\beta$ decay. Such a simple decay pattern to the doubly magic
$^{54}$Ca makes it a fundamental and clean benchmark for ab~initio methods to
test the quenching mechanism arising from 2BCs in a nuclear system with a
significant neutron excess. Across the nuclear chart, there are only a few
similar cases, such as the decay from $^{78}$Co ($Z=27$ and $N=51$) to $^{78}$Ni
($Z=28$ and $N=50$)~\cite{Tani2019} and from $^{132}$In ($Z=49$ and $N=83$) to
$^{132}$Sn ($Z=50$ and $N=82$)~\cite{Jone2010}, which are, however, beyond the
current experimental and theoretical capabilities.

Studying the $\beta$ decay of $^{54}$K is challenging both experimentally and
theoretically. On the nuclear chart, neutron-rich nuclei residing to the
southeast of $^{54}$Ca have low production rates at current radioactive ion-beam
facilities. Additionally, the $\nu f_{5/2}\rightarrow\pi f_{7/2}$ transition
populates highly excited, neutron-unbound states in $^{54}$Ca, which are only
accessible via neutron spectroscopy. Earlier attempts to measure the decay
strength of potassium isotopes reached mass number $A=53$~\cite{Perr2006,
Xu2024, Gott2025}, but for heavier isotopes, the production yield was too low to
determine anything but the half-life~\cite{Lang1983, Zeng2025}. Theoretically,
predicting the decay using ab~initio methods is complicated, requiring accurate
computations of both the ground state of $^{54}$K and the highly excited states
in $^{54}$Ca. To capture the Gamow-Teller strength to these high-lying states,
which are proton excitations from the $sd$ shell into the $pf$ shell, methods
must account for the complex many-body correlations involving multiple major
shells.

Despite these challenges, this combined experimental and theoretical study is
now possible. The Facility for Rare Isotope Beams (FRIB) makes the detailed
decay spectroscopy study of this exotic nucleus feasible. By combining
$\gamma$-ray and neutron time-of-flight (TOF) spectrometers at the FRIB Decay
Station initiator (FDSi)~\cite{FDSi, Allm2025}, we observed that the decay of
$^{54}$K mostly populated a single neutron-unbound state in $^{54}$Ca. This
measurement is compared with new ab~initio and shell-model computations, which
employ multi-shell valence spaces necessary to accurately describe the decay.
Together, our measurement and calculations demonstrate the importance of 2BCs in
$\beta$ decays in neutron-rich isotopes and support the persistence of the
$N=34$ shell closure in potassium.

\section*{Results}
\subsection*{Experiment}
The experiment was carried out at FRIB, where a selenium-82 primary beam was
accelerated up to 215 MeV/nucleon before bombarding a 5-mm carbon target with a
beam intensity of 567 particle nanoamperes (equivalent to $3.5\times10^{12}$
particles per second). The isotopes of interest were selected by the Advanced
Rare Isotope Separator (ARIS)~\cite{Haus2013, Port2023} and identified on an
event-by-event basis using the time-of-flight (TOF) versus energy loss ($\Delta
E$) method. The TOF was measured between a parallel-plate avalanche
counter in ARIS and a fast plastic scintillator in the first diagnostic cross of
the FDSi. The $\Delta E$ was taken from a silicon PIN detector located in the
same diagnostic cross. About 4600 $^{54}$K nuclei were identified over 98 hours
of beam time.

The cocktail beams containing $^{54}$K were stopped in a segmented Yttrium
Orthosilicate (YSO) implantation detector \cite{Yoko2019} at the center of the
decay station. The position and timing information were recorded for both the
implanted ions and their subsequent $\beta$ decays. The spatial and temporal
correlations allowed for a clean construction of the $\beta$-delayed spectra of
$^{54}$K with low background. Surrounding the implantation detector, there were
11 high-purity germanium clover detectors of the DEcay Germanium Array initiator
(DEGAi) and 88 modules of the Versatile Array of Neutron Detectors at Low Energy
(VANDLE)~\cite{Pete2016}. This hybrid detector system provides about 70\%
average efficiency for $\beta$ detection, 4\% full-energy photopeak efficiency
for 1-MeV $\gamma$ rays, and 20\% efficiency for 1-MeV neutrons. More
information about the FDSi setup can be found in Refs.~\citenumns{Craw2022,
Gray2023, Lubn2023, Cox2024, Ogun2025, Ong2026}.

\begin{figure}[ht]
   \centering
   \includegraphics[width=0.49\textwidth]{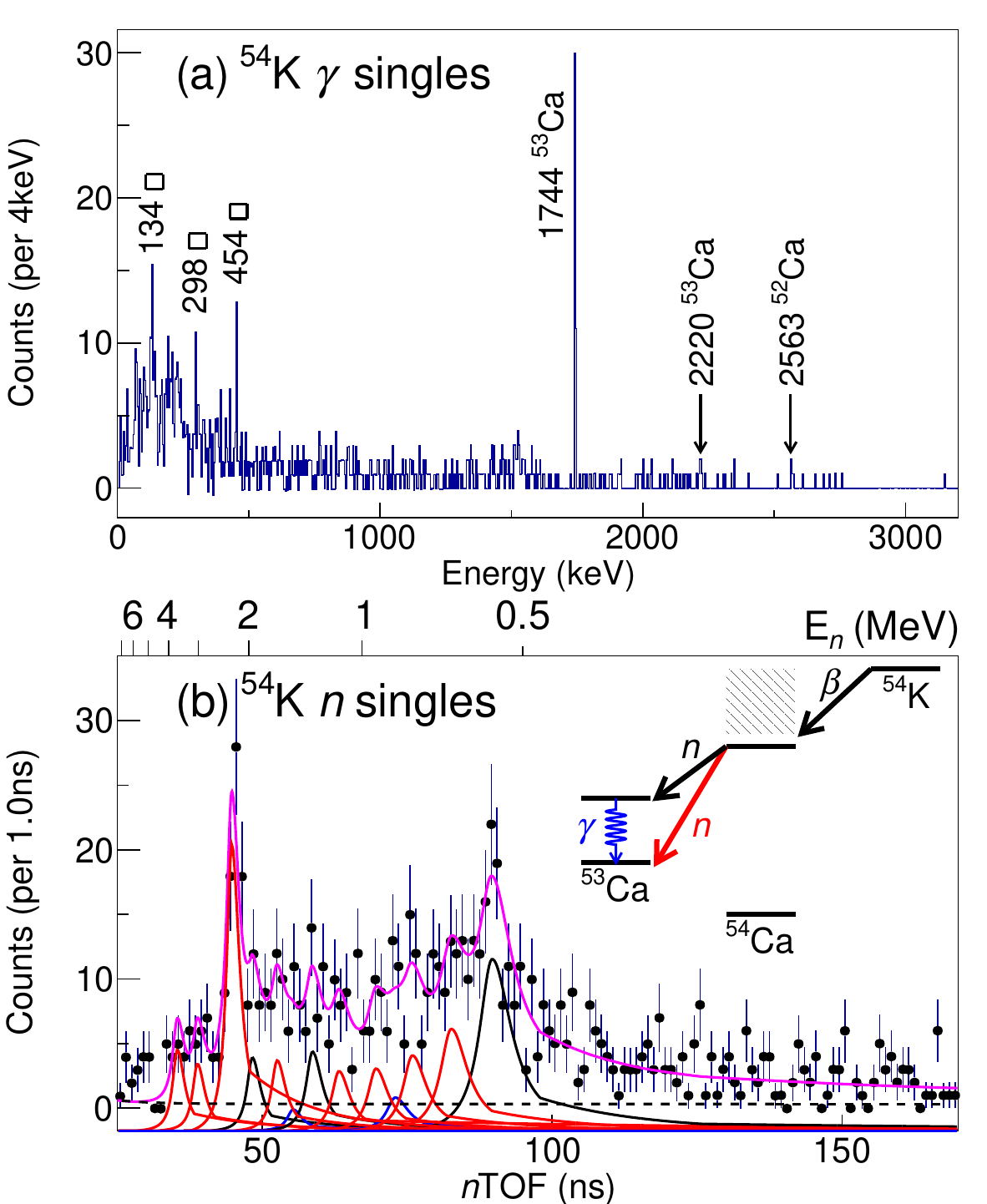}
   \caption{
      The background subtracted $\beta$-delayed $\gamma$-ray (a) and neutron TOF
      (b) spectra of $^{54}$K taken within 70 milliseconds after the ion
      implantation. The error bars represent the $1\sigma$ uncertainty. Known
      $\gamma$ transitions are labeled with their energies and origins. The
      energies followed by open squares have not been reported previously, but
      their assignments cannot be made due to the limited statistics. The TOF
      spectrum is fitted by the neutron response functions (magenta), which is
      deconvoluted into individual neutron peaks feeding the ground (red), first
      (black), and second excited state (blue) in $^{53}$Ca, respectively. The
      inset of panel (b) depicts schematically the strong decay path
      $^{54}$K$\rightarrow^{54}$Ca$\rightarrow^{53}$Ca observed in the
      experiment. The shadow box in $^{54}$Ca represents the high-lying
      unresolved states that might decay via two-neutron emissions.
   }\label{K54_exp}
\end{figure}

\Cref{K54_exp} presents the $\beta$-delayed $\gamma$-ray and neutron TOF spectra
of $^{54}$K with a 70-ms time window between the implanted ions and emitted
$\beta$ particles. The contribution from the random background was sampled using
the $\beta$ events detected preceding the $^{54}$K implantation, and was
subtracted from the histograms. In \cref{K54_exp}(a), the prominent $\gamma$
peak at 1744~keV corresponds to the $5/2^{-}\rightarrow1/2^{-}_\mathrm{\gs}$
($\gs$ stands for the ground state) transition in $^{53}$Ca~\cite{Step2013,
Chen2019}, suggesting a dominant $\beta$-delayed one-neutron emission
($\beta1n$) channel of $^{54}$K. In addition, the feeding to the $3/2^{-}$
second excited state in $^{53}$Ca is observed at 2220 keV, with a much lower
intensity compared to the $5/2^{-}$ first excited state. The population of the
first $2^+$ state in $^{52}$Ca at 2563 keV was marginal but still visible with 4
counts, giving a $P_{2n}\ge5\%$ as the lower limit of $\beta2n$ probability of
$^{54}$K. On the other hand, we could not observe the $\gamma$ ray deexciting
the first $2^+$ state in $^{54}$Ca at 2021 keV, indicating a negligible
$\beta0n$ channel. Below 1~MeV, there are a few peaks, marked by open squares in
\cref{K54_exp}(a), whose energies were not reported previously. Owing to the
limited statistics, it is not possible to see them in a $\gamma$-$\gamma$ or
neutron-$\gamma$ coincidence spectrum, preventing us from placing them as new
transitions in $^{52-54}$Ca.

The neutron TOF spectrum in \cref{K54_exp}(b) features two pronounced peaks,
between which the energy difference is about 1750 keV. Consequently, it is
reasonable to assign them as two different decay paths from the same
neutron-unbound state in $^{54}$Ca, one to the $1/2^-$ ground state and another
to the 1744-keV first excited state of $^{53}$Ca; see the inset of
\cref{K54_exp}(b). With the technique from Refs.~\citenumns{Cox2024, Xu2024},
the TOF spectrum was deconvoluted into individual neutron emissions from
$^{54}$Ca to the ground, 1744-keV, and 2220-keV ($J^{\pi}=3/2^-$) states of
$^{53}$Ca. The resulting peaks are plotted in red, black, and blue,
respectively, in \cref{K54_exp}(b). The energies and intensities of the two
aforementioned main neutron peaks reveal a strong $\beta$ feeding at 6159(100)
keV in $^{54}$Ca. Since no evidence of a competing $\gamma$-decay branch from
the state was observed, we obtained the feeding probability as
$I_{\beta}=59(8)\%$. With the Fermi function $f$~\cite{Orea1950} computed from
the AME2020 atomic mass evaluation~\cite{Wang2021} and the updated $^{54}$K
half-life of $T_{1/2}=13.3(6)$ ms from this experiment~\cite{Ong2026}, we deduce
a comparative partial half-life of $\logft=4.4(1)$ for the transition,
indicative of a Gamow-Teller decay~\cite{Sing1998}. Although the low statistics
prevent identifying the $\beta2n$ neutrons in the deconvolution, we estimated
their induced uncertainties in the fitting procedure and included them in the
uncertainty of the $\logft$ value. The impact of these uncertainties is small
because the two prominent $\beta1n$ peaks are well isolated in the spectrum.

\begin{figure}[ht]
   \centering
   \includegraphics[width=0.49\textwidth]{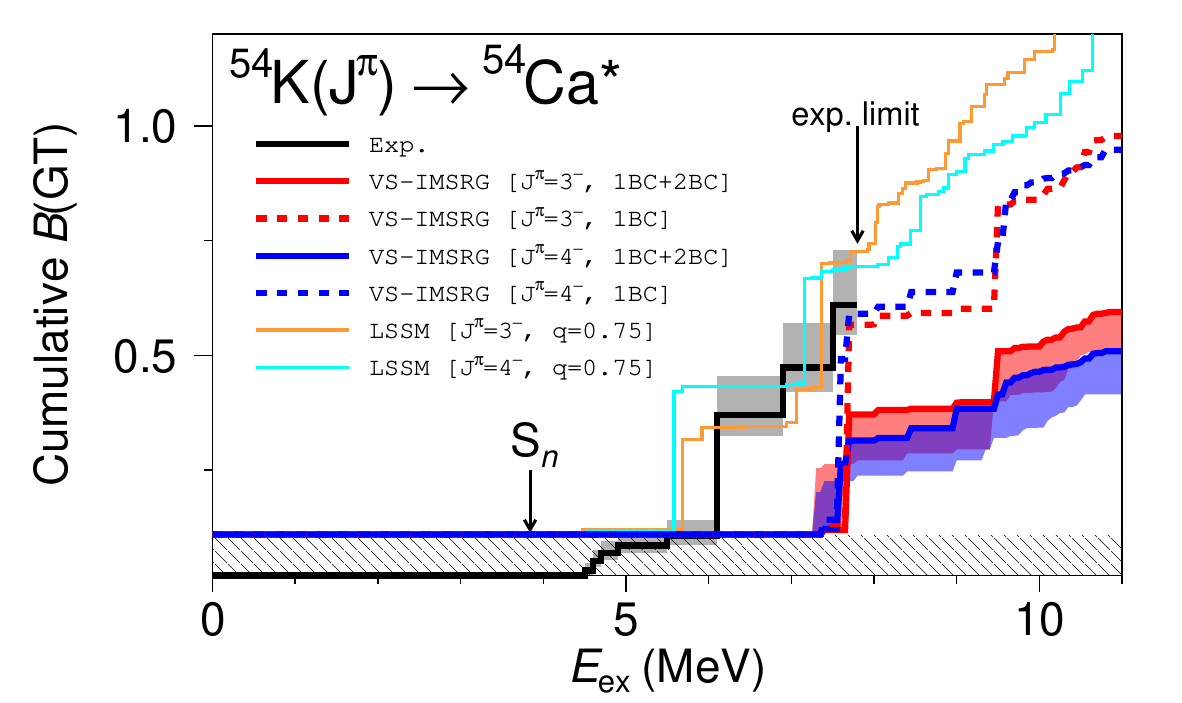 }
   \caption{
      Cumulative $\bgt$ of the $^{54}$K decay. The experimental data, which are
      plotted in the solid black line, stop around 7.8 MeV which is the upper
      limit of excitation energy in this analysis. The associated color band
      represents the $1\sigma$ uncertainty. The theoretical results are based on
      an assumed ground state of $^{54}$K with spin and parity $J^\pi$. The band
      in the VS-IMSRG calculations corresponds to the uncertainty associated
      with the multi-shell decoupling. The VS-IMSRG calculations are shown with
      and without 2BCs to illustrate their impact. While the VS-IMSRG
      calculations use the bare axial-vector coupling $g_A=1.27$, those of LSSM
      with the sdpf-mu interaction quenched the coupling by a factor $q=0.75$.
      Because the experimental strength below 6~MeV might be biased (see text
      for more details), the horizontal baselines of all the calculations were
      shifted up by 0.11, indicated by the shaded area, to compare with the data
      from 6~MeV onward.
   }\label{K54_theory}
\end{figure}

From the $\logft$ value, the Gamow-Teller transition strength can be derived as
$\bgt=K/(ft)$, with the constant
$K\equiv2\pi^3\hbar^7\mathrm{ln}2/m_e^5c^4G_\mathrm{F}^2=6144(2)$ s determined
from superallowed $\beta$ decay~\cite{Suho2007, Hard2009, Yosh2018}. Note this
definition of $\bgt$ factors in the square of axial-vector coupling $g_A^2$.
The cumulative $B(\mathrm{GT})$ distribution of $^{54}$K from the neutron TOF
deconvolution is plotted in \cref{K54_theory}, where the prominent $\beta$
feeding is reflected as a big step with $\bgt=0.26^{+0.08}_{-0.04}$ at 6159 keV.
The full list of states and their relevant information is presented in
Supplementary Table~I.

Below we infer the ground-state spin and parity of $^{54}$K. Browne \textit{et
al.}\ recently studied proton removal reactions from $^{55}$Sc
($J^{\pi}_\mathrm{\gs}=7/2^-$) to $^{54}$Ca, observing a 6166(27)-keV level
populated with a large cross section $\sigma=1.22(16)$ mb~\cite{Brow2021}.
According to their theory predictions, the states that were strongly populated
between 5 and 7 MeV in $^{54}$Ca originate from the removal of a $d_{3/2}$
proton from the $^{55}$Sc ground state. In Fig.~3 of Ref.~\citenumns{Brow2021},
three candidates of the 6166-keV state can be found in theory within the energy
range of interest: a $5^-$ state at 5.4~MeV with $\sigma=0.94$~mb, a $4^-$ state
at 5.9~MeV with $\sigma=1.35$~mb, and a $3^-$ state at 6.7~MeV with
$\sigma=0.74$~mb. Although not definitive, a spin-parity assignment other than
$(3,4,5)^-$ is much less likely given the measured energy and cross section. Our
own shell-model calculations show that those $^{54}$Ca states, with a
spin-parity of $J^-$ ($J=3$, 4, or 5) and significant proton-removal cross
sections (from the $^{55}$Sc ground state) between 5 and 7 MeV, can also be
populated in the $^{54}$K Gamow-Teller decay with a strong probability (a small
log$ft$ value) if the $^{54}$K ground state is $(J-1)^-$. In addition, as
discussed below, when comparing the newly measured $\bgt$ and theoretical
predictions, calculations starting from a $3^-$ or $4^-$ ground state of
$^{54}$K are compatible with data while those starting from a $2^-$ ground state
are not. Therefore, we suggest that the same 6166-keV state, of which
$J^{\pi}=(4,5)^-$, was observed at 6159~keV in $^{54}$Ca in our $\beta$-decay
experiment, and the ground-state spin and parity of $^{54}$K favor the
assignment of $(3,4)^-$.

\subsection*{Theory}
The experimental data are compared with theoretical predictions in
\cref{K54_theory}. We note that all the calculations feature a sizable step much
stronger than adjacent transitions between 5 and 8~MeV, suggeting it is the
theoretical counterpart of the Gamow-Teller decay feeding the 6159-keV state
observed experimentally. However, the fragmented strengths below the main decay
in experiment were not predicted by any theory. The discrepancy is attributed
to 1) the misidentified $\beta2n$ neutrons as $\beta1n$ neutrons and 2) missing
the detection of the $^{53}$Ca $\gamma$ rays in the neutron emissions feeding
the $^{53}$Ca excited states in the analysis. This implies the excitation
energies of the $\beta$ feedings below 6~MeV are likely underestimated, and they
eventually represent decay strengths to higher excited states including those
above the two-neutron separation energy in $^{54}$Ca at
7.37~MeV~\cite{Wang2021}. Since our primary focus here is the main transition at
6~MeV, for a clear demonstration of the performance of different theoretical
predictions, we shift the horizontal baselines of all the calculations up by
0.11 in \cref{K54_theory}, indicated by the shaded area, aligning them to the
experimental cumulative $\bgt$ at 6~MeV.

The ab~initio valence-space in-medium similarity renormalization group
(VS-IMSRG) calculations~\cite{Herg2016, Stro2019} (see Methods for details) of
the $\beta$ decay of $^{54}$K use two- and three-nucleon interactions and
consistent one- (1BCs) and two-body currents from chiral effective field
theory~\cite{Hebe2011, Gysb2019}. We employ a multi-shell valence space, as
shown in \cref{chart} (inset), to capture the $sd$-to-$pf$-shell correlations
necessary to describe the decay. Multi-shell calculations must control for
center-of-mass effects~\cite{Gloe1974, Miya2020}, and we quantify the
uncertainty associated with this treatment, shown as red and blue bands in
\cref{K54_theory}.

In our VS-IMSRG calculations, we cannot definitively predict the ground-state
spin of $^{54}$K: the lowest $J^\pi = 2^-$, $3^-$, and $4^-$ states are nearly
degenerate. Based on this we compute the full Gamow-Teller strength for all
three choices of parent state to the first 80 states in $^{54}$Ca, reaching
excitation energies up to nearly $11$~MeV. The calculated strength starting from
a $2^-$ state does not show any strong single transition (see Supplementary
Fig.\ 1), which is not compatible with the experiment. In contrast, computations
assuming a $3^-$ ($4^-$) initial state predicts a decay channel with sizeable
$\bgt=0.25\ (0.14)$ and 0.43 (0.35) with and without 2BCs, respectively. The
agreement in $\bgt$ is excellent for the $3^-$ state with 2BCs, whereas that for
the $4^-$ decay is moderate, agreeing within $3\sigma$ uncertainty. All computed
VS-IMSRG $\bgt$ values can be found in Supplementary Table~II and in
Ref.~\citenumns{Xu2026}. To further constrain the parent ground-state spin, we
performed ab~initio coupled-cluster calculations (see Methods for more details).
The $J^\pi=4^-$ state is the lowest in energy, followed by the $3^-$ state.
Thus, the ab~initio results for the spectra and strength function disfavor
$J^\pi=2^-$ for the spin and parity of the $^{54}$K ground state.

The solid and dashed red lines in \cref{K54_theory} clearly demonstrate that
2BCs significantly reduce the predicted Gamow-Teller strength, naturally
providing the phenomenological quenching ($\sim50\%$) that needs to be included
in our shell-model calculations described below. With 2BCs included, our
predictions of $\bgt$ agree with the experiment without any extra quenching.
However, the predicted excitation energies are systematically higher. For
example, the initial step in the cumulative $\bgt$ occurs at $7.4$~MeV,
considerably larger than what we measured. This is due to the VS-IMSRG(2)
approximation used in this work, which overpredicts excitation energies. A
recent study showed that improved VS-IMSRG calculations systematically bring the
excitation energies in neutron-rich calcium isotopes down by as much as
20\%~\cite{Hein2025} (see Methods for more details). On the other hand, previous
studies have shown that VS-IMSRG(2) calculations meet benchmarks with other
methods for the Gamow-Teller matrix elements within 3--4\% for single-state
dominated decays~\cite{Gysb2019}, suggesting improved VS-IMSRG calculations
would change the $\bgt$ relatively little. This is why the predicted cumulative
$\bgt$ are stretched out horizontally relative to the experiment. Other effects
on the calculated $\bgt$, such as that from the chosen model space, are
discussed in detail in Supplementary Information, and the results can be seen in
Supplementary Figs.\ 2--4.

Finally, we study the structures of both $^{54}$K and $^{54}$Ca with large-scale
shell-model (LSSM) calculations. We employed the sdpf-mu effective
interaction~\cite{Utsu2012} and limited the number of nucleon excitations from
the $sd$ shell to the $pf$ shell to be less than or equal to one
($\le1\hbar\omega$), which has been shown to be appropriate for $\beta$-decay
computations in this mass region~\cite{Yosh2018, Cox2024, Ong2026}. Near
$^{54}$Ca, however, this configuration produced some noticeable discrepancies
with the experimental data in low-lying states, including the excitation energy
of the $5/2^-$ state in $^{53}$Ca ($E_\mathrm{5/2^-}^\mathrm{th}=2.5$~MeV vs.\
$E_\mathrm{5/2^-}^\mathrm{exp}=1.7$~MeV~\cite{Step2013, Chen2019}), its
neutron-removal spectroscopic factor from the $^{54}$Ca ground state
($C^2S^\mathrm{th}=0.10$ vs.\ $C^2S^\mathrm{exp}=0.24$~\cite{Chen2019}), and the
excitation energy of the first $2^+$ state in $^{54}$Ca
($E_\mathrm{2^+}^\mathrm{th}=2.6$~MeV vs.\
$E_\mathrm{2^+}^\mathrm{exp}=2.0$~MeV~\cite{Step2013, Brow2021}). These
discrepancies can be improved simultaneously via lowering the $f_{5/2}$
single-particle orbital for both proton and neutron by 700~keV, giving
$E_\mathrm{5/2^-}^\mathrm{th}=1.8$~MeV, $C^2S^\mathrm{th}=0.20$, and
$E_\mathrm{2^+}^\mathrm{th}=2.0$~MeV. Adopting this modification, we then
calculate the $\bgt$ distribution of $^{54}$K assuming a ground-state spin and
parity of $2^-$, $3^-$, and $4^-$ following the method of
Ref.~\citenumns{Yosh2018}. To compare with the experimental data, a
phenomenological quenching of $q=0.75$ is applied to the axial-vector
coupling~\cite{Mart1996}, corresponding to a quenching of $q^2\approx0.56$ to
the calculated strength.

The results are consistent with the ab~initio calculations presented above:
Although LSSM predicts almost degenerate $2^-$, $3^-$, and $4^-$ states at low
energy in $^{54}$K, the Gamow-Teller strengths from either the $3^-$ or $4^-$
state in \cref{K54_theory} show much better agreement with the experimental
data than those from the $2^-$ state (see Supplementary Information),
corroborating our $^{54}$K ground-state spin assignment, $J^{\pi}=(3,4)^-$. In
addition, the good agreement between the experimental data and model predictions
allows us to investigate the $^{54}$K ground-state structure invoking
shell-model wavefunctions. The neutron occupation number on $\nu f_{5/2}$, which
is above the $N=34$ shell gap, is about 1.3 in both the $3^-$ and $4^-$ states
of $^{54}$K. Given that $^{54}$K has $N=35$ neutrons in total, its ground state
is dominated by the so-called one-particle-zero-hole configuration relative to
the $N=34$ closed shell, with a small amount of 0.3 neutrons being excited to
$\nu f_{5/2}$ from the shell below. Although slightly greater than $^{54}$Ca,
the cross-shell excitation in the $^{54}$K ground state is in stark contrast to
the neutron-rich nuclei above $^{54}$Ca, such as $^{56}$Ti, where the ground
states are dominated by the neutron excitation across $N=34$~\cite{Iimu2023}. It
suggests that the $N=34$ shell closure persists in $^{54}$K below $Z=20$, which
agrees with previous studies in this region~\cite{Liu2019, Zeng2025}.

\section*{Discussion}
\begin{figure}[ht]
   \centering
   \includegraphics[width=0.40\textwidth]{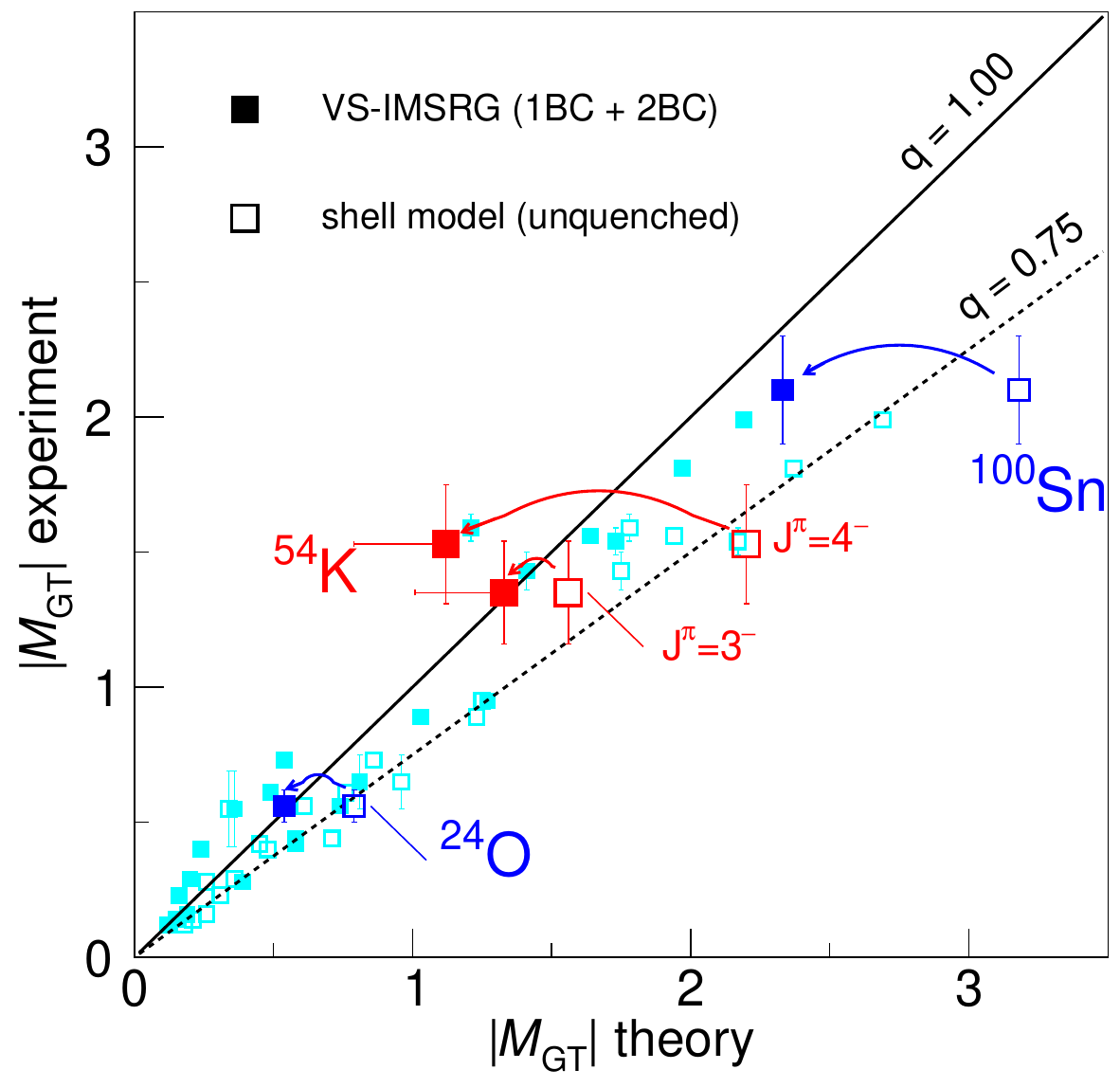}
   \caption{
      Gamow-Teller matrix elements from experimental data compared with model
      predictions. The literature data (blue for $^{24}$O and $^{100}$Sn, and
      turquoise for the rest) are taken from Refs.~\citenumns{Gysb2019,
      Neup2024}. The open and solid squares show unquenched shell-model results
      using phenomenological interactions and VS-IMSRG results with 2BCs
      included, respectively. The curved arrows for $^{24}$O, $^{100}$Sn, and
      $^{54}$K guide the eye to link together two distinct theories, VS-IMSRG
      and unquenched phenomenological shell-model calculations, of the same
      Gamow-Teller transition. Due to the undetermined ground-state spin,
      results from both the $3^-$ and $4^-$ decay of $^{54}$K are presented (red
      colors). The horizontal uncertainties are associated with the multi-shell
      decoupling. In this plot, $^{54}$K is the only nucleus of which the
      calculations employ a multi-shell valence space because of its large
      neutron excess; see \cref{chart}.
   }\label{bgt_sys}
\end{figure}

Using the results from VS-IMSRG and LSSM calculations, we extended the survey in
Refs.\ \citenumns{Gysb2019, Stro2021} to neutron-rich nuclei $^{24}$O and
$^{54}$K, comparing the experimental and theoretical Gamow-Teller matrix
elements $|M_\mathrm{GT}| = \sqrt{(2J+1)\bgt}$, with $J$ the initial spin of
$\beta$ decay. The results are plotted in \cref{bgt_sys}. For $^{24}$O, we took
the Gamow-Teller transition to the first $1^+$ state at 1830
keV~\cite{Neup2024}. In this survey, $^{54}$K is the first and only case where
such a computation is done within a multi-shell valence space; see \cref{chart}.
\Cref{bgt_sys} clearly shows that the matrix elements from LSSM with empirical
interactions (open squares) need to be quenched on average by $q\sim0.75$ to be
consistent with the experimental values. With (without) 2BCs included in the
Gamow-Teller operator, VS-IMSRG gives $|M_\mathrm{GT}| = 1.33$ (1.74) and 1.13
(1.78) for the $3^-$ and $4^-$ decay of $^{54}$K, respectively. The experimental
values of 1.35 (assuming a $3^-$ decay) or 1.53 (assuming a $4^-$ decay) are
reasonably consistent with the calculations including 2BCs (solid squares). This
illustrates that also in this very neutron-rich isotope, two-body currents
quench the matrix element significantly, giving good agreement with experiment
without requiring phenomenological adjustments.

We report on two major experimental and theoretical findings. The $\beta$-decay
strength of $^{54}$K to excited states in doubly magic $^{54}$Ca was measured
using $\beta$-delayed $\gamma$-ray and neutron spectroscopy at the FDSi. A
strong, concentrated $\beta$-decay feeding of a neutron-unbound state in
$^{54}$Ca was observed at 6159 keV, with $\logft=4.4(1)$. New VS-IMSRG
calculations using a complex multi-shell valence space reproduce the measured
Gamow-Teller strength, with quenching stemming from 2BCs. The role of 2BCs is
now demonstrated in a very neutron-rich nucleus, a case prototypical for
thousands of other decays in known but also yet unexplored parts of the nuclear
chart.

We observe that 2BCs quench the cumulative $\bgt$ almost uniformly as a function
of excitation energy. The excellent agreement between experiment and theory in
the $^{54}$K decay demonstrates the rapid progress in ab~initio computations.
Such computations now provide a consistent global description of medium-mass
nuclei rooted in nuclear forces and fundamental weak interactions. Moreover,
they are capable of capturing many-body correlations, e.g., the cross-shell
excitations for high-lying states, to reliably describe weak transitions in
exotic nuclei. Ab~initio computations are steadily paving their way towards a
global description of $\beta$ decays of open-shell, neutron-rich nuclei across
the nuclear chart~\cite{Gysb2019, Stro2021, Li2025}, and this work establishes a
key milestone along that path.

With FRIB ramping up its beam power gradually, more exotic isotopes will be
delivered for experimental study. Of particular interest are those southeast of
neutron-rich, doubly magic $^{78}$Ni and $^{132}$Sn on the nuclear chart, where
decay properties reveal emergent phenomena, and have direct impacts on the path,
time scale, and final abundances of nucleosynthesis. There, Gamow-Teller $\beta$
decay is always accompanied by one- or multi-neutron emissions together with
$\gamma$ rays. Our experiment, which simultaneously measured a variety of
$\beta$-delayed radiations from $^{54}$K with high efficiency, set the future
standard of decay spectroscopy with low production yield. The high-lying
Gamow-Teller strength addressed in those neutron-rich nuclei will continuously
challenge nuclear ab~initio theories at the limit of stability.

\section*{Methods}
\subsection*{Ab~initio calculations}
Ab~initio nuclear structure calculations solve the many-body Schrödinger
equation for nuclear Hamiltonians
\begin{equation}
    H = T_\mathrm{int} + V_\mathrm{NN} + V_\mathrm{3N}
\end{equation}
consisting of the intrinsic kinetic energy $T_\mathrm{int}$ and two- and
three-nucleon potentials $V_\mathrm{NN}$ and $V_\mathrm{3N}$ from chiral
effective field theory \cite{Epel2009, Mach2011}. We employ the 1.8/2.0~(EM)
Hamiltonian~\cite{Hebe2011}, fit exclusively to nucleon-nucleon scattering data,
the binding energy of $^3$H, and the point-proton radius of $^4$He. It has been
established to provide accurate predictions for ground-state
energies~\cite{Simo2017, Hebe2023}, excitation spectra~\cite{Hage2016,
Morr2018}, and $\beta$-decay transition strengths~\cite{Gysb2019} across the
nuclear chart.

We compute $\beta$-decay transition strengths with the valence-space in-medium
similarity renormalization group~\cite{Stro2019}. The VS-IMSRG solves the
many-body problem through a tailored unitary transformation of the Hamiltonian,
decoupling a core and a valence space from the rest of the Hilbert space. This
yields effective valence-space interactions and operators that are solved using
shell-model techniques. We perform all calculations in the VS-IMSRG(2)
approximation, where the transformation is truncated at the normal-ordered
two-body level.

To describe the $\beta$ decay of $^{54}$K, we use a multi-shell valence space to
capture the Gamow-Teller transition from the $pf$-shell neutrons to highly
excited states in $^{54}$Ca, with protons excited from the $sd$ shell into the
$pf$ shell. Specifically, we consider a $^{34}$Si core and a valence space
consisting of the neutron $pf$ shell and the proton $\pi s_{1/2}$, $\pi
d_{3/2}$, $\pi f_{7/2}$, and $\pi p_{3/2}$ orbitals. VS-IMSRG calculations with
multi-shell valence spaces are computationally intensive and must control for
center-of-mass contamination effects. We use the established Gloeckner-Lawson
prescription~\cite{Gloe1974, Miya2020} to remove and diagnose center-of-mass
contamination, using a energy denominator shift $\Delta = 10$~MeV and
center-of-mass scaling factors $\beta = 1$, $3$, $5$. We find decoupling of the
center of mass with $\beta=3$ and quantify the remaining uncertainty related to
center-of-mass contaminations by comparing results with $\beta = 3$ and $5$. We
attempted calculations with larger valence spaces, notably with a complete
proton $sdpf$ valence space. We found that in these cases the VS-IMSRG
calculations diverged, failing to decouple the larger valence space. Divergent
calculations as well as significant uncontrolled dependence on $\beta$ are two
a~posteriori diagnostics for problems with the multi-shell
decoupling~\cite{Miya2020}. The valence space employed in our calculations is at
the edge of what is currently feasible.

Our VS-IMSRG calculations are performed in a basis of 13 major harmonic
oscillator shells (with oscillator frequency $\hbar\omega = 16$~MeV) starting
from a $^{54}$K Hartree-Fock reference state. Three-nucleon forces are
additionally truncated at $E_\mathrm{3max}=24$~\cite{Miya2022, Miya2023},
sufficient for nuclei up to mass number $A\sim130$. We compute the spectra of
$^{54}$K and $^{54}$Ca, and we evaluate the Gamow-Teller transition strengths
from the lowest $2^-$, $3^-$, and $4^-$ states in $^{54}$K to the first 80
states in $^{54}$Ca. For this, we include 2BCs, which are normal ordered with
respect to $^{54}$Ca and truncated at the normal-ordered one-body level.

The ground state of $^{54}$K, as an odd-odd nucleus, is difficult to accurately
predict. Our VS-IMSRG calculations predict that the $2^-$, $3^-$, and $4^-$
states are degenerate within less than 500~keV. At $\beta=3$, the ground state
is predicted to be the $4^-$ state with the $3^-$ only 50~keV higher. Due to the
near degeneracy, we are unable to definitively predict the ground-state
spin-parity of $^{54}$K, so we consider each option as a possible parent state
for the $\beta$ decay of $^{54}$K. We find that the $\beta$ decay strengths from
the $2^-$ are substantially too small to reproduce the experimental data. This
is consistent with our coupled-cluster calculations, so we exclude this
spin-parity from further consideration.

We find that the $\bgt$ predictions for both $3^-$ and $4^-$ states are
consistent within many-body uncertainties (indicated by the band in
\cref{K54_theory}). Both predictions give similar cumulative $\bgt$ strength as
the experiment, but at higher excitation energies. This is due to the
VS-IMSRG(2) approximation. In improved calculations (which are not possible for
Gamow-Teller $\beta$ decays currently), spectra of neutron-rich calcium isotopes
are compressed by as much as 20\%~\cite{Hein2021, Hein2025}. Accounting for
this, the predicted and observed $\bgt$ strengths are consistent once 2BCs are
included. Notably, the strengths without including 2BCs are significantly too
large. This highlights the importance of 2BCs in quenching $\beta$-decay
strengths in ab~initio calculations.

For comparison with our VS-IMSRG calculations, we also perform coupled-cluster
calculations of the ground-state of $^{54}$Ca for different possible spin-parity
assignments $J^\pi$. Our calculations are performed in the coupled-cluster with
singles and doubles approximation via charge exchange~\cite{Ekst2014} from a
$^{54}$Ca ground state, computed in a basis of 13 major oscillator shells with
oscillator spacing $\hbar\omega = 16$~MeV. We find the level ordering to be
$J^\pi=4^-$ (at 0~MeV),  $3^-$ (at 0.37~MeV), $2^-$ (at 0.73~MeV) and $1^-$ (at
1.87~MeV). These results are in good agreement with those from the VS-IMSRG.

\section*{Data availability}
The calculated results presented in \cref{K54_theory,bgt_sys} can be found at
Zenodo \cite{Xu2026}. The experimental data supporting the findings are
collected from experiment E21069 at FRIB. The raw data files are extremely large
in size and contain unpublished results for other isotopes. The data reduction
procedure requires detailed instruction and guidance. They are available from
the corresponding authors upon request.

\section*{Acknowledgments}
The authors thank the FRIB accelerator staff for their efforts in delivering the
stable and high-intensity $^{82}$Se beam and in the success of this experiment.
The diagonalization of nuclear Hamiltonians and the computation of the
transition matrix elements in both VS-IMSRG and LSSM calculations are
performed using \textsc{kshell}~\cite{Shim2019}.

\section*{Funding}
This material is based upon work supported by the U.S. Department of Energy,
Office of Science, Office of Nuclear Physics, and used resources of the
Facility for Rare Isotope Beams (FRIB) Operations, which is a DOE Office of
Science User Facility under Award Number DE-SC0023633.
This work was supported by the U.S.\ Department of Energy, Office of Science,
Office of Nuclear Physics, under Award No.\ DE-FG02-96ER40983 (UTK), No.\
DE-AC05-00OR22725 (ORNL), No.\ DE-AC02-06CH11357 (ANL), No.\
DE-AC02-05CH11231 (LBNL), No.\ DE-SC0016988339
(TTU), and No.\ DE-FG02-94ER40834 (Maryland);
by the auspices of the U.S.\ Department of Energy by Lawrence Livermore
National Laboratory under Contract No.\ DE-AC52-07NA27344 (LLNL);
by the U.S.\ National Science Foundation under Grants No.\ PHY-23-10078, No.\
PHY-2209429, and No.\ PHY-2110365;
by the National Nuclear Security Administration under Award No.\ DE-NA0003180
and No.\ DE-NA0003996;
by the Stewardship Science Academic Alliances program through DOE Awards
No.\ DOE-DE-NA0003906, No.\ DOE-DE-NA0004068, and No.\ DOE-DE-NA0004074;
by the Natural Sciences and Engineering Research Council of Canada under
Grant No.\ SAPIN-2022-00019;
by the National Science Foundation Major Research Instrumentation Program
Award No.\ 1919735 (UTK, TTU);
by the Laboratory Directed Research and Development Program at Oak Ridge
National Laboratory (ORNL) and Lawrence Livermore National Laboratory (Grant
No.\ 23-LW-25);
by the Laboratory Directed Research and Development Program at Pacific
Northwest National Laboratory operated by Battelle for the U.S.\ Department
of Energy;
by the U.S.\ Department of Energy, Office of Science, Office of Advanced
Scientific Computing Research and Office of Nuclear Physics, Scientific
Discovery through Advanced Computing (SciDAC) program (SciDAC-5 NUCLEI);
by the Innovative and Novel Computational Impact on Theory and Experiment
(INCITE) program;
by the Polish National Science Centre under Contract No.\
UMO-2023/51/D/ST2/02816;
and by the Japan Society for the Promotion of Science, Grants-in-Aid for
Scientific Research (KAKENHI) under Grant No.\ 22K14053.
This research used resources of the Oak Ridge Leadership Computing Facility
at the Oak Ridge National Laboratory, which is supported by the Advanced
Scientific Computing Research programs in the Office of Science of the U.S.\
Department of Energy under Contract No.\ DE-AC05-00OR22725.
TRIUMF receives federal funding via a contribution agreement with the
National Research Council of Canada.

\section*{Author contribution}
R.G., W.J.O., J.M.A., B.C.R., K.P.R., H.S., B.M.S., and O.B.T. were the
spokepersons for the experiment. The experimental design and installation was
contributed by Z.Y.X., R.G., W.J.O., J.M.A., B.C.R., K.P.R., H.S., J.M.C., I.C.,
T.J.G., J.H., T.T.K., K.K, S.N.L., R.S.L., M.M., T.R., and D.P.S. Supplemental
equipment was provided by R.G., J.M.A., M.P.C., H.L.C., J.T.H., K.K., M.M., and
V.T. The data acquisition system was maintained by G.C., A.C., T.T.K., and N.K.
The beam transportation was operated by B.M.S., O.B.T., M.Hausmann, E.K., M.P.,
and M.K.S. The experiment at the FDSi was operated by Z.Y.X., R.G., W.J.O.,
J.M.A., B.C.R, K.P.R., H.S., S.A., H.A., A.D.A., H.C.B., J.M.B., D.L.B., K.B.,
N.B., M.P.C., J.M.C., I.C., H.L.C., B.P.C., J.D., A.A.D., J.G.D., A.E., A.F.,
C.F., K.F., T.G., T.J.G., E.C.G., K.H., S.H., J.T.H., C.M.H., A.C.H., K.H.,
D.E.M.H., D.H., J.H., R.J., M.K., T.T.K., N.K., K.K., S.N.L., B.L., R.S.L.,
S.L., M.M., M.J.M., S.N., T.H.O., G.O, J.R.P., M.M.R., A.L.R., I.R., E.K.R.,
G.E.R., T.R., N.N.S., D.P.S., D.S., K.S., A.Spyrou, M.S., A.Sweet, V.T., A.T.,
S.U., W.B.W., S.W., M.W., and R.Y. The offline analysis was contributed by
Z.Y.X., R.G., N.B., I.C., and S.N. The theoretical calculations were done by
Z.Y.X., M.Heinz, T.P., P.G., G.H., and P.N. The manuscript was written by
Z.Y.X., R.G., M.Heinz, and T.P. All authors have read and commented on the
manuscript.

\section*{Competing interests}
The authors declare no competing interests.

\end{document}